\documentstyle[12pt]{article}
\begin{document}
\rightline{SU-ITP-92-1}
\rightline{\today}
 \newcommand{\Psl}{\not\!\! P}
\newcommand{\dsl}{\not\! \partial}
\newcommand{\half}{\frac{1}{2}}
\newcommand{\for}{\frac{1}{4}}
\def\a{\alpha}
\def\b{\beta}
\def\al{\aleph}
\def\g{\gamma}\def\G{\Gamma}
\def\d{\delta}\def\D{\Delta}
\def\e{\epsilon}
\def\et{\eta}
\def\z{\zeta}
\def\t{\theta}\def\T{\Theta}
\def\l{\lambda}\def\L{\Lambda}
\def\m{\mu}
\def\f{\phi}\def\F{\Phi}
\def\n{\nu}
\def\i{\iota}
\def\p{\psi}\def\P{\Psi}
\def\r{\rho}
\def\s{\sigma}\def\S{\Sigma}
\def\ta{\tau}
\def\x{\chi}
\def\o{\omega}\def\O{\Omega}
\def\lagr{{\cal L}}
\def\cd{{\cal D}}
\def\k{\kappa}
\def\tz{\tilde z}
\def\tF{\tilde F}
\def\ri {\rightarrow}
\def\cf{{\cal F}}
\def\pa {\partial}
\def\U {\Upsilon}
\vskip 3.5 cm
\begin{center}
{\large\bf  SUPERSYMMETRIC BLACK HOLES }
\vskip 1.7 cm
{\bf Renata Kallosh \footnote{On leave of
absence from: Lebedev Physical Institute, Moscow 117924,
USSR }} \vskip 0.2cm
Physics Department, Stanford University, Stanford   CA
94305 \footnote{Bitnet address KALLOSH@SLACVM}

\end{center}
\vskip 1.6  cm
\begin{center}
{\bf ABSTRACT}
\end{center}
\begin{quote}

\ \ \ \ \  The effective action of $N=2$,   $d=4$ supergravity is
shown to acquire no quantum  corrections
in background metrics admitting super-covariantly constant
spinors. In particular, these metrics  include the
Robinson-Bertotti metric (product of two 2-dimensional spaces
of constant curvature) with  all 8 supersymmetries unbroken.
Another example is  a set of arbitrary number of extreme
Reissner-Nordstr\"om
   black
holes. These black holes break 4 of 8 supersymmetries, leaving the
other 4 unbroken.

\ \ \ \ \  \ We have found manifestly supersymmetric black
holes, which are non-trivial solutions  of the flatness
condition $ \cd^{2}  = 0$ of the corresponding (shortened)
superspace. Their bosonic part describes a set of  extreme
Reissner-Nordstr\"om     black holes. The super black hole
solutions are  exact  even when all quantum supergravity
corrections
 are taken into
 account.

\end{quote}


\newpage
\section{Introduction}
Despite all successes in quantum gravity in dimensions  $d=2,
3$,  development of this theory in  $d\geq 4$ is still a
problem. One of the main difficulties is the uncontrollable
accumulation of  divergent quantum corrections in each new
order of perturbation theory.

The purpose of this paper is to find some results in $d=4$
quantum gravity, which remain valid  with an account taken of
all orders of perturbation theory. With this purpose we
investigate some  effective quantum actions of 4-dimensional
supergravity theories in very specific backgrounds,  which
admit supercovariantly constant spinors \cite{GH}, \cite{T}.

The main result of our investigation is rather surprising: The
effective action of $d=4$  $N=2$  supergravity has no quantum
corrections in the background of arbitrary number of extreme
Reissner-Nordstr\"om     (${\cal RN}$) black holes in neutral
equilibrium. In some sense, to be defined later, the  manifestly
supersymmetric version of the extreme  ${\cal RN}$  black
holes  provides an  alternative to the trivial flat superspace.

We will start with  a discussion of some earlier results on the
vanishing of quantum corrections to  the effective actions
in supergravity theories. In $N=1, 2, 3$ supergravities the
locally  supersymmetric on shell effective
action is given in terms of the following chiral superfields:
\begin{eqnarray} \label{1}
W_{ABC} (x, \t_i ),\qquad \bar {W}_{A'  B' C'}  (x, \bar \t_i ),
\qquad  i = 1 \nonumber \\
W_{AB} (x, \t_i ),\qquad \bar {W}_{ A'  B' }  (x, \bar \t_i ),
\qquad  i = 1, 2 \nonumber\\ W_{A} (x, \t_i ),\qquad \bar
{W}_{ A'  }  (x, \bar \t_i ), \qquad  i = 1, 2, 3
\end{eqnarray}
Consider for example quantum corrections to $N=1$
supergravity \cite{RK},  \cite{GG}. The superfields are
\begin{eqnarray} \label{2}
W_{ABC} (x, \t)&=&\Psi_{ABC}(x)  + C_{ABCD} \t^D + ...
\nonumber\\
\bar {W}_{A'  B' C'}(x, \t)&=&\bar {\Psi}_{A'B'C'}(x) +
\bar {C}_{A'B'C'D'} \bar {\t}^{D'} + ... \ ,
\end{eqnarray}
where $\P, \bar \P $ are the gravitino field strength spinors
and  $C, \bar C $ are the Weyl spinors of the space-time. Each
locally supersymmetric term in the effective  action depends
both on $W$ and $\bar W$ and  their covariant derivatives.
\footnote{The only exceptions are the one-loop topological
divergences proportional  to $W^2$ or $\bar W^2$, related to
the one-loop anomalies.} For example, the troublesome 3-loop
counterterm is \cite{DS}  \begin{equation}
S_{3-loop} = \int d^4x \,d^4 \t \, \det E    \quad
W_{ABC}W^{ABC}  \bar {W}_{A'  B' C'}  \bar {W}^{A'  B' C'} \ .
\end{equation}
It is a supersymmetrized square of the
Bell-Robinson tensor. This term, as well as any other term of
the effective quantum  action, vanishes in the {\it
super-self-dual} background {\cite{RK}}
 \begin{equation}
W_{ABC} = 0\ ,  \qquad      \bar {W}_{A'  B' C'}\neq 0\ ,
\end{equation}
or in the {\it super-anti-self-dual} background
\begin{equation}
W_{ABC} \neq 0\ ,  \qquad      \bar {W}_{A'  B' C'} = 0\ .
\end{equation}
Such non-trivial backgrounds exist only in space-time with
Euclidean  signature. Indeed,  in Minkowski space
\begin{eqnarray} \label{6}
W_{ABC} &\Rightarrow  & W^{real} + i W^{im}\ , \nonumber\\
\bar {W}_{A'  B' C'} &\Rightarrow  & W^{real} - i W^{im}  \ .
\end{eqnarray}
Therefore $W$ and $\bar W$ cannot vanish separately, only
together, in which case the background is trivial.
With Euclidean signature it is possible to have a vanishing
right-handed spinor $W_{ABC} (x, \t)$ and a non-vanishing
left-handed spinor $\bar {W}_{A'  B' C'}(x, \bar \t) $
(or opposite). This {\it half-flat superspace}, where the
left-handed gravitino $\bar {\Psi}_{A'B'C'}(x)$
lives in the space with only left-handed curvature
 $\bar {C}_{A'B'C'D'}(x) $, is the background where $N=1$
supergravity effective action has no quantum corrections
(up to the above-mentioned topological terms).
\vskip 0.5 cm
In $N=2$ supergravity we have
\begin{equation}\label{7}
W_{AB}(x, \t) = F_{AB}(x) + \Psi_{ABC}^{i}(x)\, \t_{i}^C   +
C_{ABCD}\, \t^C_i \t^D_j \e ^{ij} + ... \ ,
\end{equation}
where $F_{AB} $ is the Maxwell field strength spinor. In the
Euclidean half-flat superspace, where
\begin{equation}
W_{AB} = 0\ ,  \qquad      \bar {W}_{A'  B'} \neq 0\ ,
\end{equation}
there are no quantum corrections to the effective action (up to
topological terms).
For $N=3$ the half-flat superspace is given by $W_{A} = 0  \, ,
\bar {W}_{A' } \neq 0$.

To summarize,  some examples of non-trivial background field
configurations in
supergravity, which receive no radiative corrections, have
been known  for more than 10 years  \cite{RK}, \cite{GG}.
They all require
Euclidean signature of space-time.

\section{Absence of Quantum Corrections in Robinson-Bertotti
background}  The special role of the Robinson-Bertotti metric
in the context of the solitons in supergravity was explained in
lectures by  Gibbons  \cite{GH}. His proposal was to consider
the Robinson-Bertotti (${\cal RB}$)  metric as an alternative,
maximally supersymmetric, vacuum state. The extreme
Reissner-Nordstr\"om     metric spatially interpolates between this
vacuum and the trivial flat one, as one expects from a soliton.

In what follows we are going to prove a non-renormalization
theorem for the effective action of $d=4, N=2$
 supergravity in the ${{\cal RB}}$ background.

The ${ {\cal RB}}$ metric is known to be one particular
example of a  class of
metrics, admitting  super-covariantly  constant spinors
\cite{GH}, \cite{T}, which are called Israel-Wilson-Perjes
(${\cal IWP}$) metrics \cite{KRAM}. It is also the special
metric in this class, which does not break any of the 8
supersymmetries of $d=4, N=2$ supergravity; all other ${\cal
IWP}$  metrics break at least half of the supersymmetries.

In general relativity the ${ {\cal RB}}$ metric is known as the
conformally flat solution of the Einstein-Maxwell system with
anisotropic electromagnetic field
 \cite{KRAM}. It describes the product of two 2-dimensional
spaces of constant curvature:
\begin{equation}
ds^2 = \frac{2d\zeta d\bar \zeta}{[1+ \a \zeta \bar \zeta]^2} -
\frac{2du dv}{[1+ \a uv]^2 }\ ,  \qquad  \a = const\ .
\end{equation}
The metric can also be written in the form
\begin{equation}
ds^2 = (1 - \l y^2) dx^2  +  (1 - \l y^2)^{-1} dy^2  +    (1 +\l
z^2)^{-1} dz^2  -   (1 + \l z^2) dt^2 \ .
\end{equation}
The corresponding Maxwell field $F_{ab}$ is constant (as well
as the curvature tensor) and can be written as
\begin{eqnarray} \label{11}
F_{12}^{{\cal RB}} &=&  \sqrt  {2\l }\ \sin \b \ , \nonumber\\
F_{34}^{{\cal RB}} &=&  \sqrt  {2\l }\ \cos \b \ , \nonumber\\
\l & = & const\ , \nonumber\\
\b & = & const\ .
\end{eqnarray}
The property of the ${\cal RB}$ metric which is of crucial
importance  for our analysis is the conformal flatness of this
metric, i.e. the vanishing of the Weyl tensor.
\begin{equation}\label{13}
C_{abcd}^{{\cal RB}} = 0 \quad \Rightarrow
C_{ABCD}^{{\cal RB}}
=  C_{A'B'C'D'}^{{\cal RB}}  = 0 \ .
\end{equation}
The curvature spinor, corresponding to the traceless Ricci
tensor,  satisfies Einstein's equation
\begin{equation}
R_{A'B' AB}^{{\cal RB}}  = F_{AB}^{{\cal RB}}
\bar F_{A'B'}^{{\cal RB}} \ .
\end{equation}
The generic term in the effective quantum action of $N=2$
 supergravity is given by
\begin{equation} \label{15}
 \G \sim \int d^4x \,d^{4N} \t  \, \det E \   A ( W_{AB},
\bar {W}_{A'  B' }, D_{CC'} W ,  D_{CC'} \bar W , \dots ) \ ,
\end{equation}
where $A$ can either be a local or non-local function  in $x$
of  the superfields $W , \,\bar W$ and their covariant
derivatives, the superfield $W$ being given in eq. (\ref{7}).

For  trivial flat  superspace $W= \bar W = 0$, since there are
no Maxwell, gravitino or Weyl curvatures in the flat
superspace. Therefore the path integral of $d=4, \, N=2$
supergravity has no quantum corrections in the flat
superspace.

Consider now the superfield $W$, containing   $F_{AB}^{{\cal
RB}} $ of the  ${\cal RB}$ solution, given in  eq.  (\ref{11}), and
recall that there is no gravitino nor Weyl spinors in this
background. The superfield $W, \, \bar W$ is a constant
superfield but it does not vanish as it would be the case for the
trivial flat superspace. It has only the first component in the
expansion in $\t $ i.e. it does not dependend on  $\t $  at all.
\begin{equation}\label{16} W_{AB}^{{\cal RB}} = F_{AB}^{{\cal
RB}},  \quad  \bar W_{A'B'}^{{\cal RB}} =
\bar F_{A'B'}^{{\cal RB}}
\ . \end{equation}
 Now we only have to look for the terms in eq.
(\ref{15}) which depend on $W , \,\bar W$  but not on their
covariant derivatives in bosonic or fermionic directions, since
these derivatives are zero for the  ${\cal RB}$ solution:
\begin{equation} \label{17}
 \int d^4x\, d^{4N} \t  \det E\,  \ A ( W_{AB}^{{\cal RB}} ,
\bar {W}_{A'  B' }^{{\cal RB}} ) \  .
\end{equation}
If $A$ is a local function of $W , \,\bar W$, this expression
takes  the form
 \begin{equation} \label{18}
A ( W_{AB}^{{\cal RB}} ,  \bar {W}_{A'  B' }^{{\cal RB}} )
\int d^4x d^{4N} \t \det E = A \,\int dV = 0    \ ,
\end{equation}
since the invariant volume of the real superspace vanishes
\cite{ES} in $N=2$ supergravity. For
non-local functions we end up with the integral over the
volume of  the full superspace of certain functions of $x$.
Those integrals are also equal to zero, according to \cite{ES}.

Thus, we have proved that there are no quantum corrections to
the  effective action of $N$\,=\,2 supergravity in the
Robinson-Bertotti background. The basic difference with
trivially flat superspace  is the fact that the  superfield, in
terms of which the on shell quantum corrections are
expressed, is not zero, but is a constant superfield. However,
all supersymmetric invariants vanish as in the case of a flat
superspace with vanishing superfield. \vskip 0.5 cm
\section{Absence of Quantum Corrections in the Extreme
Black Hole
Background}

The proof of the non-renormalization theorem for the  ${\cal
RB}$  background was almost trivial due to conformal flatness
of this metric and because the Maxwell field is constant.
These properties are not present for  general metrics admitting
super-covariantly constant spinors.  In general relativity they
are known as conformal-stationary class of Einstein-Maxwell
fields with conformally  flat 3-dimensional space. This class
of metrics has been found by Neugebauer, Perjes, Israel and
Wilson \cite{KRAM}:
\begin{eqnarray}\label{19}
ds^2 &=& (V\bar V )^{-1} (dt + {\bf A} d{\bf x})^2 -
 (V\bar V )\,
(d{\bf x} )^2  \quad  \nonumber \\
{\bf \nabla} \times {\bf A} &=& - i (\bar {V} {\bf \nabla} V -
\bar {V} {\bf \nabla} V ), \quad {\bf \nabla}^2  V = 0,
\quad V \neq
0 \ ,
\end{eqnarray}
where ${\bf \nabla}^2 $ is the flat space Laplacian in ${\bf x}
$. For real $V$ this metric reduces to the
Majumdar-Papapetrou  solutions \cite{KRAM}, which,
according to  Hartle and Hawking \cite{HH}, are the only
regular black hole solutions in this class. They describe an
arbitrary number $n$ of extreme ${\cal RN}$ black holes with
gravitational attraction balanced by electrostatic repulsion:
 \begin{equation}\label{20}
V= \bar V = 1 + \sum_{s=1}^{s=n} \frac{M_s}{|{\bf x} - {\bf x}
_s|}\ . \end{equation}

It has been found by Gibbons and Hull \cite{GH}, and in the
most  general form by Tod \cite{T}, that these metrics
admit super-covariantly constant spinors of $N=2$
supergravity.  We will reformulate here the results of
\cite{GH}, \cite{T} for the
special case of pure $N=2$ supergravity.\footnote{
 Tod's parameter $Q$, related to dust density is equal to
zero in our theory since there is no dust in pure supergravity.}

We have found that in the treatment of super-covariantly
constant  spinors of $N=2$ supergravity it is very helpful to
use the original Penrose notation \cite{PR}. We introduce a
standard spinor basis, or dyad,  $o^A,\, \i^A$,   and we define
\footnote{Our spinorial indices take values 0,1 and $0', 1'$.
The Greek letters chosen by Penrose for the basis:
$o, \, \i $ (omicron and iota) visually resemble
these numbers.
 The use of the equations  $\e_0^A =
o^A, \, \e_1^A = \i ^A$ and many others is
particularly simple in this
notation.}

\begin{equation}\label{21}
\e_{AB} \ o^A \i ^B = o_A \i ^A = V,    \quad  \e_{A'B'} \ o^{A'}
\i ^{B'}  = o_{A'} \ \i ^{A'} = \bar V \ ,
\end{equation}

Only when $V=\bar V = 1$ the dyad is a spin frame. However,
it  is possible to work with a dyad which is  not normalized to
unity. Associated with any spinor basis of the manifold is a
null tetrad $l^a, n^a, m^a, \bar m^a$ defined by
\begin{eqnarray}\label{22} l^a = o^A o^{A'},\quad n^a = \i^A
\i^{A'},\quad  m^a=o^A \i^{A'}, \quad \bar m^a=\i ^A o^{A'} \ ,
\end{eqnarray}
and satisfying the following conditions:
\begin{eqnarray}\label{23}
l^a l_a = n^a n_a = m^a m_a = \bar m^a \bar m_a &=& 0 \ ,
\nonumber\\
l^a m_a = l^a \bar m_a = n^a m_a = n^a \bar m_a&=& 0 \ ,
\nonumber\\
l^a n_a =- m^a \bar m_a &=& V\bar V \ .
\end{eqnarray}

We will require that these spinors are super-covariantly
constant  \cite{GH}, \cite{T}.
\begin{eqnarray}\label{24}
\nabla _{AA'} \  o_B + F_{AB} \  \i _{A'} &=& 0 \ ,
\nonumber\\  \nabla _{AA'} \  \i _{B'} - F_{A'B'}\  o _{A} &=&
0 \ .
\end{eqnarray}

{}From now on we will limit ourselves to the case of black holes
only  (real  $V$), postponing the treatment of more
 general metrics to a future publication. In particular,
basically  the same techniques can be used to study eq.
(\ref{21}) with complex $V$,
 i.e. when ${\bf \nabla} \times {\bf A}$ in eq.  (\ref{19})
in the definition of the ${\cal IWP}$ metric is non-vanishing.
Other interesting examples  of metrics
admitting super-covariantly constant spinors are plane-wave
space-times
 for which $V=0$ in eq.  (\ref{19}). They also will be
considered  in a separate publication.

\vskip 0.5 cm
  We use that
\begin{equation}\label{25}
 o^A  = V^{-1} K_{AA'} \ \i ^{A'}, \quad \i _{A'} = -
V^{-1} K_{AA'} \ o^A\ ,
\end{equation}
where $K_a$ is the Killing vector
\begin{equation}\label{26}
K_{AA'} = (l+n)_{AA'}\ .
\end{equation}

Eqs. (\ref{24}) can be rewritten as follows
\begin{eqnarray}\label{27}
\triangle _{AA'}\  o_B  &\equiv & \hat {\nabla}_{AA'}\  o_B
-   V^{-1} K_{AA'} F_{BC} \ o^C = 0 \ , \nonumber\\
\triangle _{AA'} \ \i _{B'} &\equiv &  \hat {\nabla}_{AA'}\
\i _{B'} -  V^{-1} K_{AA'} F_{B'C'}\  \i^{C'} = 0 \ .
\end{eqnarray}
The hatted derivatives have the standard meaning \cite{PR} of
derivatives in the conformally rescaled metric
\begin{eqnarray}\label{28}
\hat {g}_{ab} = V^2 g_{ab}\ ,  \quad \U_a = V^{-1}\nabla_a V \ ,
\nonumber\\
 \hat {\nabla}_{AA'}\ o_B  = \nabla _{AA'}\ o_B -  \U _{BA'}\
o_A \ . \end{eqnarray}

If the null tetrads are expressed according to eq. (\ref{22})
through super-covariantly  constant omicron and iota
satisfying   eqs. (\ref{24}),
 one gets the following equations for differential
forms \footnote{The wedge product symbol is omitted for
simplicity, when multiplying forms.}
\begin{eqnarray}\label{29} \hat {d}m \equiv dm &=&  0 \ ,
\nonumber\\ \hat {d}\bar m \equiv  d\bar m &=&0 \ ,
\nonumber\\ \hat {d}(l-n) \equiv  d (l-n) &=& 0 \ ,
\nonumber\\ \hat {d}K  \equiv dK - 2 \U K &=& 0  \ ,
\nonumber\\  \hat {d} A \equiv dA -  \U A &=& 0\ ,
\nonumber\\  \hat {d} \U \equiv d\U &= &0 \ ,\nonumber\\
\hat {d} w^{ab} - w^{ac} w_c^{\ b} \equiv   d w^{ab}  -
w^{ac} w_c^{\ b}  &= &0 \ ,
\end{eqnarray}
where  the null tetrad and Maxwell curvature forms are
defined as
  \begin{equation} \label{30}
 m= dx^a m_a,\quad  \bar m= dx^a \bar m_a,\quad  l= dx^a l_a,
\quad  n= dx^a n_a,\quad K = l+n, \quad F = dA \ ,
\end{equation}
and we have  introduced the Lorentz connection form $w^{ab}$
and  the Weyl connection form $\U$.

 These equations can be solved as follows:
  \begin{eqnarray}\label{31}
m&=& 2^{-\half}\, (dx+i dy)\ , \nonumber\\
\bar m&=& 2^{-\half}\, (dx-i dy)\ , \nonumber\\
 l-n &=&\sqrt 2 \, dz \ , \nonumber\\
K &=& \sqrt 2  \,V^2  dt   \ , \nonumber\\
A &=& V dt\ ,\nonumber\\
\U &=& V^{-1}dV \ ,\nonumber\\
w_{AB} &=&V^{-1}( \i_A d o_B  - o_A d\i_B  +
\half  \e_{AB}\, dV )\ .
\end{eqnarray}
This leads to the Papapetrou-Majumdar metrics
 \begin{equation} \label{32}
 ds^2 = V^2 dt^2 - V^{-2} d{\bf x}^2\ ,  \quad F  = dV  \wedge
dt\ ,
 \end{equation}
 where the flat-space Laplacian in $x,y,z$ of $V$ is zero
 and $F$ may still be subject to some dual rotation. These
coordinates $x,\,y,\,z$ are called comoving
coordinates and $t$ is defined as  $K^a\nabla_a =
\sqrt 2 \,\frac{\partial }{\partial t}$.

To calculate the curvature of the manifold we act with
$\nabla_{CC'}$ on equations (\ref{24}).
 \begin{eqnarray} \label{33}
 R&=&0\ , \nonumber\\
 R_{A'B'AB} &=& F_{AB}\, \bar {F}_{A'B'}\ , \nonumber\\
 C_{ABCD} &=& \nabla_{AB'} \, F_{CD} \, V^{-1}\,  K^{B'}_{\ B} \ ,
\nonumber\\
\bar C_{A'B'C'D'} &=& \nabla_{A'B} \, \bar {F}_{C'D'} \, V^{-1} \,
K^{B}_{\ B'}\ .
\end{eqnarray}

Now we have enough information to investigate the effective
quantum action in the black hole background \footnote {The
equations presented above for black holes can be  derived also
for the ${\cal RB}$ metric. There will be a second set of
covariantly constant spinors, defined by eqs.
 (\ref{24}) with opposite sign in front of $F$ and $\bar F$ for
the  second set of omicron and iota. The second Killing vector
will be built from the second set of these spinors.  Both $C^+$
and $C^-$  are equal to zero for ${\cal RB}$
 as a consequence of all those equations.}
 with the properties:
 \begin{eqnarray}\label{34}
F_{AB}(x) &=& V^{-2}\, K_A^{\ A'}\,  \nabla_{A'B} \, V\ ,
\nonumber\\
\P_{ABC}^{i}(x) &=& 0\ , \nonumber\\
 C_{ABCD}^-  &\equiv &C_{ABCD} -  \nabla_{AB'}\,  F_{CD}\,
V^{-1} \,
K^{B'}_{\ B} = 0 \ ,\nonumber\\
C_{ABCD}^+  &\equiv &C_{ABCD} +  \nabla_{AB'} \, F_{CD} \,
V^{-1} \,
K^{B'}_{\ B} \neq 0 \ ,
\end{eqnarray}
and the conjugate ones can be easily derived from eqs.
(\ref{34}).

The basic on shell superfield $W$ in the real basis is given
by
\begin {equation}
W_{AB}(x, \t, \bar \t) = F_{AB}(x) + \Psi_{ABC}^{i}(x)\
\t_{i}^C   +
 C_{ABCD}\  \t^C_i\, \t^D_j \, \e ^{ij} +
\nabla _{C D'} F_{AB}\ \t^C_i \,\bar \t ^{D' i} + \dots  \ .
\end{equation}
The first component of this superfield is neither zero, as in
flat  superspace, nor a constant, as  in ${\cal RB}$ case.  The
second component is zero, we have just a bosonic background.
To analyze the second component we first have to change
variables . Instead of working with independent unconstrained
8 fermionic coordinates  $\t^A_i , \, \bar \t ^{B' i}$ of the real
$N=2$ superspace, for the black holes we need the following 16
coordinates, satisfying 8 constraints:
 \begin{eqnarray} \label {36}
\t^{ \pm}_{Ai} & \equiv& \t_{A i}  \pm E_{A A'} \, \e_{ji} \,
\bar \t^{A' j}\ , \nonumber\\
\bar \t^{A' i  \pm }& \equiv& \bar \t^{A' i }
\pm  E^{ A' A }\, \e^{ij} \,  \t_{Aj} \ , \nonumber\\
\t_{Ai}^{ \pm}& =& \pm E_{A A' }\,  \e_{ji} \,  \bar \t^{A'
j\pm}\ .
\end{eqnarray}
Here $E$ is a normalized Killing vector,
\begin{eqnarray}\label{37}
E_A^{\ A' } &=& V^{-1} K_A^{\ A' } \ , \nonumber\\
E_{A A' } E^{A' B} &=& \d_A^{\ B} \ , \nonumber\\
\e_{ij} \e^{kj} &= &\d_i^{\ k} \ .
 \end{eqnarray}
In terms of these coordinates, whose supersymmetry variation
is
 \begin{eqnarray}
 \e^{ \pm}_{Ai} & \equiv &\d \t^{ \pm}_{Ai}   =  \e_{Ai}
\pm E_{\ AA' } \, \e_{ji}\, \bar \e^{A' j}\ , \nonumber\\
 \bar \e^{A' i  \pm } & \equiv &\d \bar \t^{A' i  \pm } =  \bar
\e^{A' i }  \pm E^{\ A' A}\, \e^{ij} \, \e_{Aj }\ ,
 \end{eqnarray}
the supersymmetry breaking and the shortening of the
unbroken  superspace related to extreme ${\cal RN}$ black
holes can be understood. The supersymmetric transformation
of the gravitino  field strength is
\begin {equation}\label{39}
\d \Psi_ {ABCi} = C_{ABCD}^{+} \e^{D+}_i  + C_{ABCD}^{-}
\e^{D-}_i  \ .
\end{equation}
It can be made zero under two conditions.  The first is
\begin {equation}\label{40}
 C_{ABCD}^{-}  \equiv C_{ABCD} -  \nabla_{AB'} F_{CD}
V^{-1}  K^{B'}_{\ B} = 0 \ ,
\end{equation}
which is satisfied for the black holes according to eqs.
(\ref{34}).  This condition is the property of extreme black
holes  that some combination of curvature and Maxwell fields
vanish.   It is an integrability condition for the existence of
supercovariant spinors $\e^{A-}_i $ (\ref{24}).
The second condition is
\begin {equation}\label{41}
\e^{A+}_i = \d \t_i^{A+} = 0 \ ,
\end{equation}
and requires the breaking of 4 supersymmetries. It also
indicates that after  the change of coordinates, given by  eqs.
(\ref{36}), there are  4 independent combinations of fermionic
coordinates. They are given by 8 coordinates $\t^{A -}_i , \,
\bar \t^{A' i  -}$, constrained by 4 conditions   \begin
{equation}\label{42} \t^{A -}_i = - E^A_{\ A' } \e_{ji} \bar
\t^{A' j-}\ .
\end{equation}
These combinations are still unbroken coordinates of the
superspace, since $\e^{A-}_i $ in eq. (\ref{39}) can take
arbitrary values and the variation of gravitino nevertheless
vanishes.

\vskip 0.5 cm
At this point it is appropriate to explain the difference
between Robinson-Bertotti solution and black holes from the
point of view of supersymmetry. Both metrics belong
to the general class of Israel-Wilson-Perjes metrics, admitting
super-covariantly constant spinors. For ${\cal RB}$ both
combinations, $C^+$ and $C^-$, which define the
supersymmetry transformation of the gravitino field strength,
vanish,  since the gravitational Weyl tensor and the derivative
of a Maxwell  tensor vanish separately. Therefore there are no
restrictions on  supersymmetry variations of all 8 coordinates
of the superspace, i.e. both $\e^{A+}_i $  and $\e^{A-}_i $ are
arbitrary and nevertheless the supersymmetry variation of
gravitino field strength is zero in the ${\cal RB}$ background.

\vskip 0.5 cm

We have to analyze the structure of quantum corrections
before  using the properties of the black hole background.
However, we will
 work with  variables
which are natural for this problem, like Weyl-Maxwell
spinors  $C^{\pm}, \, \bar C^{\pm}$, given in eqs.  (\ref{34})
and fermionic coordinates of the superspace, given in eqs.
(\ref{36}), considering the vector $E_{AA'}$ as some arbitrary
one. When the quantum corrections are calculated in an
arbitrary Lorentz-covariant background, there is no
dependence on any such vector, of course. We have introduced
this dependence through our choice of variables, and it should
be absent in terms of the original variables after integration
over fermionic variables.

 In the black hole background all terms which depends on
$C^{-}$  or $ \bar C^{-}$ will vanish. The crucial question is:
Are there terms which depend only on $C^{+}, \, \bar C^{+}$
and do not depend either on $C^{-}$
 or on $\bar C^{-}$?  The answer is no, they do not exist.  The
point is that under the $\e^{A-}_i $ transformations the
non-vanishing combinations of curvature and Maxwell  fields
$C^{+}, \, \bar C^{+}$  do transform. However, the $\e^{A-}_i $
variation coming from any term containing fermions will have
the combinations $C^{-} \e^{C-}_i $, according to eq.
(\ref{39}). Any term with  $C^{+}$  or  $\bar C^{+}$
dependence but without $C^{-}$ or  $\bar C^{-}$ dependence
will not satisfy the $\e^{A-}_i $ -supersymmetry
requirements. To illustrate this general statement consider
again the 3-loop counterterm. The following combinations can
be expected.  \begin
{equation}\label{43} (C^+)^2 (\bar C^+)^2, \quad (C^-)^2 (\bar
C^-)^2, \quad (C^+ C^- )(\bar C^+ \bar C^- )\ , \quad etc.
\end{equation}
Only the first combination does not vanish in the
black hole background. Let us show that it will not appear in
the effective quantum action. The straightforward calculation
is to check the dependence of each of these terms on the vector
$E_{AA'}$ by substituting expressions for $C^{\pm}$ from
eqs. (\ref{34}). The term $(C^+)^2 (\bar C^+)^2 $, which is
forbidden by the above mentioned supersymmetry arguments,
 does depend on $E_{AA'}$, as opposed to the third term in
(\ref{43}), which is allowed by the  $\e^{A-}_i
$-supersymmetry and can be shown to be
$E_{AA'}$-independent.
 \begin{eqnarray}\label{44}
 (C^+ C^- )(\bar C^+ \bar C^- )&= & \{(C_{ABCD} +
\nabla _{AA'} F_{CD} E^{A'}_{\ B})
 (C^{ABCD} - \nabla ^{A}_{B'} F^{CD}E^{B'B} )\} (\bar C^+ \bar
C^-) \nonumber\\
 &=& \{ (C_{ABCD}) ^2  - (\nabla _{AA'} F_{CD})^2\}
  \{ (\bar C_{A'B'C'D'}) ^2  - (\nabla _{AA'} \bar F_{C'D'} )^2\}
 \ .
 \end{eqnarray}

Thus, all terms in the on shell effective  quantum action of
$N=2$,   $d=4 $ supergravity, which are locally
supersymmetric and Lorentz invariant, vanish in the extreme
multi black hole background.

\section{Black Holes as a Flat Superspace}

Our approach to the black hole superspace was inspired by
the  group manifold approach to $N=2$ supergravity
\cite{CAF}.  The superspace \cite{CAF} is formulated in terms
of the superspace  1-forms
$E^a, \,\P _A^i,\, \P_{A'i},\, A$ associated to the supergravity
physical fields and the spin connection $w^{ab}$. We are
interested only in on shell curvatures associated with $d=4,\,
N=2$ super-Poincar\'e
 algebra with central charge. They are defined in terms of
the following differential operator:
\begin {equation}\label{45}
 \cd = d + A^M T_M \ ,
\end{equation}
where $T_M $ are the generators of the super-Poincar\'e
group
\begin {equation}\label{46}
[T_M, \, T_N \} = f^L_{MN} T_L \ ,
\end{equation}
and $A^M $ are connection forms $E^a, \,\P _A^i,\, \P_{A'i},\,
A_{ij}, \, w^{ab}$ related to $P_a, \, Q^{A}_{ i}, \, Q^{A'i}, \,
M^{ij}  \ , \, M_{ab}$ generators of super-Poincare group
(translation,  8 supersymmetries, central charge and  Lorentz
generator). Thus, the operator $\cd$ is
\begin {equation}\label{47}
 \cd = d + E^a P_a + \P _A^iQ^A_i + \P_{A'i}Q^{A'i} +
A_{ij}M^{ij} +
 w^{ab}M_{ab}
 \ .
\end{equation}

The curvature of this superspace is defined as
\begin {eqnarray}
 \cd ^2  &=&  R^M T_M \ , \nonumber\\
 R^M &=&  d  A^M  + f^M_{LN} A^L A^N\ .
\end{eqnarray}
The set of curvatures includes
 \begin{eqnarray}\label{49}
R^a &=& dE^a - w^{ab} E_b - i \bar \P ^i \g^a \P _i   \equiv
DE^a -
 i \bar \P^i  \g^a \P _i\ , \nonumber\\
\r _{Ai} &=& d \P _{Ai} -  w^{ab} (\s _{ab} \P _i)_A
\equiv D \P _{Ai} \ ,\nonumber\\
\r _{A'}^{i} &=& d\P_{A'}^i -   w^{ab} (\s_{ab} \P^i)_{A'}
\equiv D \P _{A'}^{i}\ , \nonumber\\
F &=& dA +(\e^{AB} \e^{ij}\P _{Ai} \P _{Bj}\ + h.c.) \ ,
\nonumber\\
R^{ab} &=& dw^{ab} - w^a_c w^{cb}\ .
  \end{eqnarray}

The nilpotency of the operator $\cd$ is the requirement that
all  curvatures are equal to zero $\cd ^2  =  R^M T_M = 0$,
or in detail
 \begin{eqnarray}\label{50}
R^a &=& 0 \ , \nonumber\\
\r _{Ai} &=& 0\ ,\nonumber\\
\r _{A'}^{i} &=& 0\ , \nonumber\\
F &=& 0\ ,\nonumber\\
R^{ab} &=& 0\ .
  \end{eqnarray}

Equation (\ref{50}) has a trivial solution describing a flat
$d=4, \, N=2$  superspace with 4 bosonic, 8 fermionic
coordinates and pure fermionic central charge form:
 \begin {eqnarray}\label{51}
E^a &=& dx^a - i \sum_i \bar \t  \g^a d \t\ , \nonumber\\
\P _{Ai} &=& d \t _{Ai}\ , \nonumber\\
\P _{A'}^{i} &=& d \t _{A'}^{i}\ , \nonumber\\
A &=& \e^{AB} \,\e^{ij}\, \t _{Ai}\, d \t _{Bj}\ + h.c. \ ,
\nonumber\\ w^{ab} &=& 0 \ .
\end{eqnarray}

Now we will build a non-trivial flat superspace for the black
holes.  The variables which are natural for the black hole
superspace are  not Lorentz covariant objects, like in
eq.  (\ref{50}), but Lorentz invariant ones, like in eqs.
(\ref{29}),  (\ref{30}) , (\ref{22}). The bosonic forms
are  $ m, \     \bar m,\   l - n ,\   K = l+n$ and  $A$.
In addition we introduce 8 fermionic forms satisfying 4
constraints, in accordance with the shortening of a black  hole
superspace. These forms are required to be super-covariantly
constant:
\begin{eqnarray}\label{52}
\nabla _{AA'} \P_B^i  &+&
F_{AB} \e^{ij} \P _{A'j} = 0\ , \nonumber\\
 \P_A^i &=&- E_{A A'} \e^{ij} \P ^{A'}_{j} \ ,
\end{eqnarray}
where  $ E_A^{A'} $ is the normalized Killing vector, defined
in  eqs. (\ref {37}), (\ref {26}), (\ref {21}). We choose the
fermionic forms of the black hole superspace to be
\begin
{eqnarray}\label {53} \P_A^1 &= &o_A \p _1 + \i_A \p_2 \
,\nonumber\\ \P_{A}^{ 2} &= &o_A \p^{2 \dagger} -  \i_A \p^{1
\dagger} \ ,\nonumber\\ \P_{A'1} &= &o_{A'} \p ^{1\dagger} +
\i_{A'} \p^{2 \dagger} \ ,\nonumber\\ \P_{A'2} &=& o_{A'}
 \p_2 -
\i_{A'} \p_1 \ .
\end{eqnarray}
Omicron and iota in equations (\ref {52}) satisfy eqs.  (\ref
{24}).  Our 4 independent Lorentz invariant fermionic forms
$\p ^i, \, \p ^{i\dagger}$ satisfy the following equations:
\begin {eqnarray}
\hat d \p^i &=& d \p^i = 0 \ , \nonumber\\
\hat d \p^{i\dagger} &=&d \p^{i\dagger} = 0\ .
\end{eqnarray}
They have a simple solution:
\begin {eqnarray}
 \p^i &=& d \t^i \ , \nonumber\\
 \p^{i\dagger} &=&d\t^{i\dagger} \ .
\end{eqnarray}
Thus, our space contains 4 fermionic coordinates in addition
to the 4  bosonic ones. \vskip 0.5 cm

The differential operator $\cd$ for the black hole superspace
is
 defined as follows:
\begin {equation}\label{56}
\cd = d + KP_K +mP_m +\bar mP_{\bar m}+ (l-n) P_{l-n} +
\p^i Q_i + \p ^{i\dagger} Q_i^{\dagger} +
A_{ij}M^{ij} + w^{ab}M_{ab} + \U W \ ,
 \end{equation}
where $P_K , \, P_m, \, P_{\bar m}, \,  P_{l-n} $ are the
translation operators in the null tetrad basis. $Q_i, \, Q^{i
\dagger}$ are the 4 supersymmetry generators, $M^{ij} = \e^{ij} M,
\, M_{ab}$ are generators of central charge and  Lorentz
symmetry and by $W $ we have denoted the generator of
conformal transformations.

The manifestly supersymmetric generalization of the bosonic
multi black hole can be obtained by solving  the flatness
condition $\cd^2 \equiv  (dA^M + f^M_{NL} A^N A^L) T_M = 0$
of the black hole superspace with the set of curvatures $R^M$
required to vanish  \footnote{Our notation here corresponds to
the one in \cite{PR}. The corresponding reference with
supersymmetry, matching the notation of \cite{PR}, is
\cite{BH}.} :
\begin{eqnarray}\label{57} d m &=& 0\ ,
\nonumber\\
 d \bar m &=&0 \ , \nonumber\\
 d (l-n) &=& 0 \ , \nonumber\\
d K - 2 \U K + 2 \sqrt 2  i  V^2  \p^2 &=&0  \ , \nonumber\\
 d \p_i &=& 0 \ ,\nonumber\\
d \p^{i\dagger} &=& 0\ ,\nonumber\\
dA -  \U A +   2  i V \p^2  &=& 0\ , \nonumber\\
 d w^{ab}  - w^{ac} w_c^b  &=& 0 \ ,\nonumber\\
 d\U &=& 0 \ ,
\end{eqnarray}
where $ \p^2 \equiv  \p_i  \p^{i\dagger}$.

These equations can be solved as follows:
  \begin{eqnarray}\label{58}
m&=& 2^{-\half} (dx+i dy)\ , \nonumber\\
\bar m&=& 2^{-\half} (dx-i dy)\ , \nonumber\\
 l-n &=&\sqrt 2 dz \ , \nonumber\\
K &=& \sqrt 2  V^2  (dt  - i \t_i d\t^{i\dagger} +
i \t^{i\dagger} d \t_i  )  \ , \nonumber\\
 \p_i &=& d \t_i \ ,\nonumber\\
 \p^{i\dagger} &=&d\t^{i\dagger} \ ,\nonumber\\
A &=& V (dt - i \t_i d\t^{i\dagger} + i \t^{i\dagger} d \t_i )  \ ,
\nonumber\\
w_{AB} &=&V^{-1}(  \i_A d o _ B  - o_A d\i_B  +
\half  \e_{AB} dV ) \ ,\nonumber\\
\U &=& V^{-1}dV \ ,
\end{eqnarray}
 the flat-space Laplacian in $x,y,z$ of $V$ is zero. In
particular,  $V$ can be chosen in the form
(\ref {20}), and in this case {\it eqs. (\ref {58}) define a
manifestly supersymmetric multi black hole}: a flat
superspace with 4 bosonic and 4 fermionic coordinates, the
flatness condition being defined as the vanishing of all
curvatures    (\ref {57}) in this superspace. This superspace,
when considered at $ \t_i = \t^{i\dagger} = d \t_i  =
d\t^{i\dagger} = 0$, coincides with the set of extreme
Reissner-Nordstr\"om black holes in the usual 4-dimensional
bosonic space.

\vskip 0.5 cm
Thus, to find an extreme black hole we were  solving  not the
classical Einstein-Maxwell equations but the flatness
condition  of the superspace. The solution of the classical
Einstein-Maxwell equations defines the Ricci tensor in terms
of a quadratic combination of Maxwell tensors, the
Riemann-Christoffel curvature tensor is non-vanishing. All  on
shell quantum gravity corrections are expressed in terms of
the non-vanishing Riemann-Christoffel curvature tensor in the
case of non-extreme as well as extreme ${\cal RN}$ black
holes. In both cases it is impossible to handle quantum
corrections without additional information.

The additional information indeed exists for the extreme
${\cal RN}$ black holes. It is  possible is to investigate the
effective quantum action of $d=4, \, N=2$ supergravity, taking
into account the fundamental fact that the bosonic extreme
${\cal RN}$ black holes do not break 4 of the original 8
supersymmetries. The investigation shows that all on shell
quantum corrections, which are locally supersymmetric and
Lorentz invariant, vanish in the extreme  ${\cal RN}$ multi
black hole background.

Moreover, the 4 above mentioned supersymmetries can be made
manifest. A shortened superspace exists (\ref {56}) , whose
flatness condition (\ref {57}) is solved by a supersymmetric
black  hole solution (\ref {58}). All curvatures of this
superspace vanish, and there are no geometrical building
blocks for quantum corrections. In this sense the
supersymmetric black holes represent an alternative to the
trivial flat superspace (\ref {47}), (\ref {51}), which also has
no quantum corrections because
 all curvatures of the superspace vanish (\ref {50}).

\vskip 0.6 cm
Our investigation was greatly stimulated by the recent activity in
strings and black hole physics \cite{W}. It is a pleasure to
express my gratitude to S. Giddings, G. Gorowitz, M. Green, J.
Hartle, A. Linde, T. Ort\'in, A. Strominger,  L. Susskind and L.
Thorlacius for interesting and fruitful discussions.  \vskip 0.6 cm

This work was supported in part by NSF grant PHY-8612280
and  by the John and Claire Radway Fellowship in the School of
Humanities and Sciences at  Stanford University.

\pagebreak

\end{document}